\documentclass{article}
\usepackage[utf8]{inputenc}
\usepackage{amsmath}
\usepackage{graphicx}
\usepackage{ wasysym }
\usepackage{authblk}
\usepackage{euscript}
\usepackage{cite}
\usepackage{hyperref}
\usepackage{tikz}
\usepackage{amsfonts}
\title{M Theory from F Theory}
\author[*]{Warren Siegel}
\author[*]{Di Wang}
\affil[*]{C. N. Yang Institute for Theoretical Physics \authorcr State University of New York, Stony Brook, NY 11794-3840}
\date{\today}

\begin{document}
\maketitle
\vspace*{-9cm}
\begin{flushright}
{~\\
YITP-SB-20-34}
\end{flushright}
\vspace*{+7cm}
\begin{abstract}
We write down a $GL(D+1)$ (D for the dimension of string theory) manifest fundamental brane worldvolume current algebra description of M theory, which consists of a pair of vector field $X^m$ and dual 2-form field $X_{mn}$, compositing together to parametrize the spacetime, with a selfduality condition for sectioning. The worldvolume of the brane itself is a (D+2) dimensional object, and the background spacetime after sectioning has dimension (D+1). We summarize the features of the algebra. The field contents of the corresponding background geometry, the usual vielbein $e_a{}^m$ and the 3-form $A_{mnp}$, could be identified as different blocks of the composite spacetime vielbein, by solving the orthogonality condition. Their behaviour under gauge transformation are also determined by the corresponding rules of the composite vielbein. Then by solving F theory $\EuScript{V}$ constraints, we reduce the number of worldvolume and show how to recover M theory from F theory.
\end{abstract}
\newpage
\tableofcontents
\newpage
\section{Introduction}
F theory, first showed up as a theory behind a particular D-7 brane background solution\cite{Vafa:1996xn}, is a generalization of string theory, with all the S($SL(2)$), T($O(D,D)$), and U($E_{D+1}$) duality symmetry being manifest. The massless sector of F theory is the exceptional field theory\cite{Coimbra:2011ky,Berman:2012vc,Godazgar:2014nqa,Musaev:2014lna,Hohm:2013pua,Hohm:2013vpa,Hohm:2013uia,Hohm:2014fxa,Hohm:2019bba} whose solution is usually called U manifold\cite{Kumar:1996zx,Berman:2020tqn}.
\\

Previously we constructed F theory, from the first quantized point of view, with a set of current algebra of branes\cite{Linch:2015fca}, as a generalization of string or T theory worldsheet description\cite{Siegel:1993bj,Siegel:1993th,Siegel:1993xq,Polacek:2013nla,Polacek:2016nry,Hatsuda:2018tcx,Hatsuda:2017tfa,Hatsuda:2015cia,Hatsuda:2014qqa,Hatsuda:2014aza}. In this approach, we started with the fundamental brane, an object different from D-brane or M-brane\cite{Hatsuda:2012vm,Hatsuda:2013dya}, whose worldvolume and spacetime coordinates are both extended to take representations in the same $E_{D+1}$ U duality symmetry group and thus tied up together. These extra stuffs could be refined by the applications of $\EuScript{S}$ ,$\EuScript{U}$, $\EuScript{V}$ section conditions. With the help of such technique, we can not only probe the background from the torsion constraint of the algebra in curved spacetime and see the exceptional field theory effectively, but also conceptually capture all the stringy excitations and corrections\cite{Linch:2015lwa,Siegel:2019wrr}.
\\

The D dimensional F theory duality symmetry group $E_{D+1}$ would be spontaneously broken into smaller pieces, if we solve these section conditions in an explicit way\cite{Polacek:2014cva}. Solving the $\EuScript{S}$ section condition or the selfduality constraint (SD for short) would separate the momentum from winding modes while solving the $\EuScript{U}$ section condition would pick up a specific worldvolume direction. Their effects and the relationship between F, M, T, and S theory duality symmetry groups could be summarized as shown below. The massless bosonic fields live in the coset and isotropy groups being symmetry groups of the vacuum, which, for F theory, follow from the general properties of spinors in dimension D mod 8\cite{Linch:2016ipx}:
\begin{center}
\begin{tabular}{ccccc}
&&F: $\frac{E_{D+1}}{H_D}$ &$\EuScript{U}\to$& T: $\frac{O(D,D)}{O(D-1,1)^2}$ \\
&&&& \\
Internal $\frac{GL(10-D)}{SO(10-D)}$& $\otimes$ & \qquad$\EuScript{S}\downarrow$ or SD & &\qquad$\EuScript{S}\downarrow$ or SD \\
&&&& \\
&&M: $\frac{GL(D+1)}{SO(D,1)}$&$\EuScript{U}\to$&S: $\frac{GL(D)}{SO(D-1,1)}$ \\
\end{tabular}
\end{center}

In this paper, we first formulate a $GL(D+1)$ manifest current algebra of M theory, including a pair of currents carrying the vector and dual 2-form representations of $GL(D+1)$ respectively. The usual massless bosonic field contents, the graviton and 3-form potential, could be read out from different blocks of the composite vielbein in this approach. The algebra itself could be constructed via the standard Legendre transformation from a Lagrangian living in an enlarged coset $GL(D+2)/SO(D,2)$. The Lagrangian contains a vector field, together with a D-1 form field living on the D+2 dimensional worldvolume of a fundamental brane and this pair of fields are related to each other by a dual condition added by hand. 
\\

Then we demonstrate how it can be revealed as a subset of F theory, by solving $\EuScript{V}$ section condition on worldvolume, with respect to the $GL(D+1)$ subgroup. The number of worldvolume would therefore be reduced and, as a result, some components of the currents would commute with everything else and thus decouple from the algebra. Finally the net remaining stuffs are the M theory vector and dual 2-form current. We show such procedure from both $\EuScript{L}$ and $\EuScript{H}$ approaches\cite{Siegel:2018puf}.
\section{M theory from fundamental brane}
We start from the action:
$$
S=\int  dX_1\wedge (*dX_1)+dX_{D-1}\wedge(*dX_{D-1})
$$
with an additional selfduality condition imposed by hand:
$$
dX_1=*dX_{D-1}
$$
where $X_1$ \& $X_{D-1}$ stands for 1-form \& (D-1)-form respectively. Note that in the Lagrangian formalism, the symmetry group gets enlarged to $GL(D+2)$.
\subsection{Current algebra}
Relabeling $X_{ijk}=X_3=*X_{D-1}$, then breaking $GL(D+2)\to GL(D+1)$, we define the conjugate momentum:
\begin{align*}
    P_m=&\eta_{mn}\dot{X}^n=-i\frac{\partial}{\partial X^m} \\
    P^{mn}=&\eta^{mp}\eta^{nq}\dot{X}_{pq0}=-i\frac{\partial}{\partial X_{mn0}}
\end{align*}
and treat $X^0$ and $X_{mnp}$ as Lagrangian multiplier. Then switching into Hamiltonian formalism, we have a set of $GL(D+1)$ covariant current algebra and the $SO(D,1)$ invariant Hamiltonian density, which can be summarized in the following way:
\begin{align*}
\EuScript{H}&=\rhd_m\rhd^m+\frac{1}{2}\rhd_{mn}\rhd^{mn} \\
\rhd_m&=P_m+\frac{1}{2}\partial^nX_{mn} \\
\rhd^{mn}&=P^{mn}+\frac{1}{2}\partial^{[m}X^{n]} \\
i[ \rhd_m(1),\rhd_n(2)]&=0\\
i[\rhd_m(1),\rhd^{np}(2)]&=\delta_m^{[n}\partial^{p]}\delta(1-2)\\
i[\rhd^{mn}(1),\rhd^{pq}(2)]&=0\\
\end{align*}
with $\EuScript{U}$ constraints, from $X^0$ \& $X_{mnp}$ variation, which generates the gauge transformation:
\begin{align*}
\EuScript{U}&=\rhd_m\partial^m=0\\
\EuScript{U}^{mnp}&=\rhd^{[mn}\partial^{p]}=0\\
\end{align*}
and the $\EuScript{S}$ constraints, which would generate worldvolume diffeomorphism up to $\EuScript{U}$ when acting on other operators, from auxiliary worldvolume vielbein variation:
\begin{align*}
    \EuScript{S}^m&=\rhd_n\rhd^{mn}=0 \\
    \EuScript{S}^{mnpq}&=\rhd^{[mn}\rhd^{pq]}=0 \\
    i[\EuScript{S}^m(1) ,\rhd_{p}(2) ]&=-\rhd_p(1)\partial^m\delta(1-2)+\delta_m^p\EuScript{U}\delta(1-2)\\
    i[\EuScript{S}^m(1),\rhd^{pq}(2)]&=-\rhd^{pq}(1)\partial^m\delta(1-2)+\EuScript{U}^{pqm}\delta(1-2)\\
    i[\EuScript{S}^{mnpq}(1),\rhd_r(2)]&= \delta_r^{[m}\EuScript{U}^{npq]}\delta(1-2)\\
     i[\EuScript{S}^{mnpq}(1),\rhd^{rs}(2)]&=0 \\
\end{align*}
and a pair of second class selfduality conditions:
\begin{align*}
    \Tilde{\rhd}_m&=P_m-\frac{1}{2}\partial^nX_{mn}=0 \\
\Tilde{\rhd}^{mn}&=P^{mn}-\frac{1}{2}\partial^{[m}X^{n]}=0 \\
\end{align*}
Solving either the selfduality condition or the $\EuScript{S}^m$ sectioning may give us the same result. We can choose the ($P_m$, $X^m$) background subspace and neglect ($P^{mn}$, $X_{mn}$), and eventually things become into:
\begin{align*}
\rhd_m&=P_m\\
\rhd^{mn}&=\partial^{[m}X^{n]}\\
\EuScript{U}&=P_m\partial^m=0 \\
\EuScript{U}^{mnp}&=\partial^{[m}X^n\partial^{p]}=0\\
\end{align*}
The $\EuScript{S}$ constraints are:
\begin{align*}
    \EuScript{S}^n&=P_m\partial^{[m}X^{n]}=-P_m\partial^nX^m \\
    \EuScript{S}^{mnpq}&=(\partial^{[m}X^n\partial^p)X^{q]} \equiv 0 \\
\end{align*}
up to the application of $\EuScript{U}$ constraint. $\EuScript{S}^{mnpq}$ becomes redundant.
\subsection{Curved background}
Coupling to curved backgrounds would be achieved by decorating vielbein fields to the currents, turning them into covariant derivatives of both spacetime and worldvolume:
\begin{align*}
\rhd_A&=E_A{}^M\rhd_M \\
D_a&=e_a{}^m\partial_m \\
\end{align*}
The spacetime vielbein, with indices written explicitly, has the following structure:
$$
E_A{}^M=
\begin{pmatrix}
E_a{}^m & E_{anp}\\
E^{bcm} & E^{bc}{}_{np} \\
\end{pmatrix}
$$
Now we can compute the current algebra in this curved background to get the expressions for torsion and constraints between different components of the vielbein from $\eta$ being an invariant tensor.
\begin{align*}
    i[\rhd_A(1),\rhd_B(2)]&=T_{AB}{}^C\rhd_C\delta(1-2)+\eta_{AB}{}^{\EuScript{A}}D_{\EuScript{A}}\delta(1-2) \\
    T_{AB}{}^C&=\nabla_{[A}E_{B]}{}^M (E^{-1})_M{}^C-\frac{1}{2}\eta_{[A|D}{}^{\EuScript{A}}\eta^{CE}{}_{\EuScript{A}}\nabla_E E_{|B]}{}^M (E^{-1})_M{}^D \\
\end{align*}
The orthogonality conditions, classified by the type of indices, are such a set of equations:
\begin{align*}
E_{(a|mn}E_{|b)}{}^{[m}e_{c}{}^{n]}&=0\\
E_a{}^mE^{bc}{}_{mn}e_d{}^n+E_{amn}E^{bcm}e_d{}^n&=\delta^{[b}_a\delta^{c]}_d \\
E^{abm}E^{cd}{}_{mn}e_e{}^n+E^{ab}{}_{mn}E^{cdm}e_e{}^n&=0\\
\end{align*}
We can write down a simple solution, with everything expressed in terms of the worldvolume vielbein $e_a{}^m$ and a 3-form field $A_{mnp}$:
\begin{align*}
E_a{}^m&=e_a{}^m\\
E_{amn}&=e_a{}^pA_{mnp}\\
E^{abm}&=0\\
E^{ab}{}_{mn}&=e_m{}^{[a}e_n{}^{b]}\\
\end{align*}
and, correspondingly:
$$
(E^{-1})_M{}^A=
\begin{pmatrix}
e_m{}^a & -e_b{}^re_c{}^sA_{mrs}\\
0 & e_{[b}{}^ne_{c]}{}^p\\
\end{pmatrix}
$$
By choosing the ($P_m$, $X^m$) pair, we set the background fields being functions of $X^m$ only, by which the algebra acting on fields would be simplified with:
$$[\rhd^{mn},E_A{}^M]=\nabla^{mn}E_A{}^M=0$$
We then list the non-zero parts of torsion:
\begin{align*}
    T_{ab}{}^c&=\nabla_{[a}e_{b]}{}^me_m{}^c\\
    T_{a}{}^{bc}{}_{de}&=\nabla_{[a|}e_m{}^{[b|}e_{|d}{}^m\delta^{|c]}_{e]} \\
    T_{abcd}&=\nabla_{[a}A_{bcd]} \\
\end{align*}
The corresponding gauge transformations of vielbeins are defined by:
$$
\delta_\lambda =i[\int \lambda^A\rhd_A,]
$$
Finding the behaviour of this set of vielbeins under gauge transformation is straight forward:
\begin{align*}
    \delta E_A{}^M\rhd_M:=\delta_\lambda( E_A{}^M\rhd_M)&=i[\int\lambda^B\rhd_B,E_A{}^M\rhd_M] \\
    \delta E_A{}^M(E^{-1})_{M}{}^B&=\lambda^CT_{CA}{}^B-\nabla_A \lambda^B +\eta_{AC}{}^\EuScript{A}\eta^{BD}{}_\EuScript{A} \nabla_D \lambda^C \\
\end{align*}
More explicitly, we have:
\begin{align*}
    \delta e_a{}^m&=\lambda^b\nabla_be_a{}^m-\nabla_a \lambda^m \\
    \delta A_{abc}&=-\nabla_{[a}\lambda_{bc]}\\
\end{align*}
which are indeed the gauge transformation rules for vielbein and 3-form in supergraivty.
\subsection{M to S(tring) theory}
Solving the $\EuScript{U}$ constraints after solving selfduality condition would break $GL(D+1)\to GL(D)$ and help us to pick up a specific worldvolume direction (labeled by "-1" or $\sigma$). Eventually, we get the usual string current algebra:
\begin{align*}
    \EuScript{U}=P_m\partial^m=0&\to\partial^{-1}=\partial_\sigma, P_{-1}=\partial^i=0\\ 
    \EuScript{U}^{[mnp]}=\partial^{[m}X^n\partial^{p]}&\to0\\
    \rhd_m&\to\rhd_{-1}=0,\rhd_i=P_i \\
    \rhd^{mn}&\to\rhd^{ij}=0,\rhd^{-1i}=\partial_\sigma X^i \\
    \EuScript{S}^m&\to\EuScript{S}^i=0,\EuScript{S}^{-1}=P_i\partial_\sigma X^i \\
\end{align*}
\section{M theory form F theory}
\subsection{Lagrangian approach}
By solving the Lagrangian version of $\EuScript{V}$ sectioning and imposing the selfduality condition for field strength, the Lagrangian we constructed could show up from the F theory Lagrangian directly. For later convenience, here we list the the enlarged symmetry group F for F theory Lagrangian field strength in each D, for details on the branching rules between different groups, see\cite{Yamatsu:2015npn}:
\begin{center}
    \begin{tabular}{ccc}
D & G                     & F \\
1 & $SL(2)$               & $SL(3)$ \\
2 & $SL(3)SL(2)$          & $SL(4)SL(2)$ \\
3 & $SL(5)$               & $SL(6)$ \\
4 & $SO(5,5)$             & $SO(6,6)$ \\
5 & $E_6$                 & $E_7$\\
\end{tabular}
\end{center}
\begin{enumerate}
    \item[$\bullet$]D=1
    \\F theory and M theory share the same Lagrangian in this case.
    \item[$\bullet$]D=2
    \\The F field strength is $F_{mni}=\partial_{[m}X_{n]i}$ and the selfduality condition is $F_{mni}=\frac{1}{2}\epsilon_{mn}{}^{pq}C_{i}{}^jF_{pqj}$. Then breaking the $SL(2)$ and identifying $X_1$ \& $X_2$ with the desired 1-form \& (D-1)-form (which is also 1-form in this case) would be good.
    \item[$\bullet$]D=3
    Now we have $\partial_m$: $6\to 5\oplus 1$, $X_{mn}$: $15\to 10\oplus5$ and $F_{mnp}$: $20=10_+\oplus10_-\to 10\oplus10$. With the scalar worldvolume derivative dropped, we have:
    \begin{align*}
        F_{10}^+&\to \partial_{[m}X_{n]}+\epsilon_{mn}{}^{pqr}\partial_pX_{qr}\\
        F_{10}^-&\to \partial_{[m}X_{n]}-\epsilon_{mn}{}^{pqr}\partial_pX_{qr}=0\\
    \end{align*}
    Thus the pair of fields and the selfduality condition could be seen clearly.
    \item[$\bullet$]D=4
    The decomposition rules of $SO(6,6)\to GL(6)$ is:
    \begin{align*}
        X^\alpha(32)\to& X_{m}(6_{-2})\oplus X_{mnp}(20_0)\oplus X^m(\bar{6}_2) \\
        \partial_m(12) \to &\partial_m(6_1)\oplus \partial^m(\bar{6}_{-1}) \\
        F_\alpha(\bar{32})\to &F_{mn}(15_{-1})\oplus F^{mn}(\bar{15}_1)\oplus F_1(1_3)\oplus F_2(1_{-3})\\
    \end{align*}
    $\EuScript{V}$ constraint is then $\eta^{mn}\partial_m\partial_n\to \partial_m\partial^m=0$, and we can choose a solution $\partial^m=0$. Then:
    \begin{align*}
        F_{mn}&=\partial_{[m}X_{n]} \\
        F^{mn}&=\epsilon^{mnpqrs}\partial_pX_{qrs}\\
        F_1&=\partial_mX^m \\
        F_2&=0\\
    \end{align*}
    Selfduality condition says $F_1=F_2=0$ and $F_{mn}=F^{mn}$ (using $SO(3,3)$ metric to move indices).
    \item[$\bullet$]D=5
    \\We list the behaviour of objects under branching $E_7\to GL(7)$:
    \begin{align*}
        X\& \EuScript{V}(133)\to& 48_0\oplus 1_0\oplus 7_8\oplus\bar{7}_{-8}\oplus 35_{-4}\oplus \bar{35}_4\\
        \partial\& F(56)\to& 7_{-6}\oplus 21_{2}\oplus \bar{21}_{-2}\oplus\bar{7}_6 \\
    \end{align*}
    Then the set of $\EuScript{V}$ constraint is:
    \begin{align*}
        \EuScript{V}_{48\oplus 1}&=\partial_7\otimes\partial_{\bar{7}}+\partial_{21}\otimes\partial_{\bar{21}}\\
        \EuScript{V}_{35}&=\partial_7\otimes\partial_{21}\\
        \EuScript{V}_{\bar{35}}&=\partial_{\bar{7}}\otimes\partial_{\bar{21}}\\
        \EuScript{V}_{\bar{7}}&=\partial_7\otimes\partial_{\bar{21}} \\
        \EuScript{V}_{7}&=\partial_{\bar{7}}\otimes\partial_{21} \\
    \end{align*}
    Although $\partial_{21}$ being non-zero could be a solution, the selfduality condition thereafter would then be too strong and everything would be eliminated. So we pick up $\partial_7=\partial_m$ non-zero and set all the others to 0. The field strength turns into:
    \begin{align*}
        F_m&=\partial_nX_m{}^n \\
        F_{mn}&=\partial_{[m}X_{n]} \\
        F^{mn}&=\epsilon^{mnpqrst}\partial_pX_{qrst} \\
        F^m&=0 \\
    \end{align*}
    with selfduality conditions $F_m=F^m=0$ and $F_{mn}=F^{mn}$.
\end{enumerate}
\subsection{Hamiltonian approach}
We can start with the Bosonic $E_n$ F-theory selfdual current algebra and its $\EuScript{S}$, $\EuScript{U}$, $\EuScript{V}$ constraints and selfduality condition:
\begin{align*}
\rhd_M&=P_M+\frac{1}{2}\eta_{MN}{}^\EuScript{I}\partial_\EuScript{I} X^N \\
    i[\rhd_M(1),\rhd_N(2)]&=\eta_{MN}{}^\EuScript{I}\partial_\EuScript{I} \delta(1-2) \\
    \EuScript{S}_\EuScript{I}&=\eta^{MN}{}_\EuScript{I}\rhd_M\rhd_N=0 \\
    \EuScript{U}_\lambda&=\EuScript{U}_\lambda^{N\EuScript{J}}\rhd_N\partial_\EuScript{J}=0\\
    \EuScript{V}_{\lambda N}&=\EuScript{U}_\lambda^{M\EuScript{I}}\eta_{MN}{}^\EuScript{J} \partial_\EuScript{I} \partial_\EuScript{J}=0 \\
    \Tilde{\rhd}_M&=P_M-\frac{1}{2}\eta_{MN}{}^\EuScript{I}\partial_\EuScript{I} X^N=0 \\
\end{align*}
We then consider the details, case by case.
\begin{enumerate}
\item[$\bullet$]D=1
\\This is actually a trivial case, as the M theory duality group is $SL(2)$, same as the one for F theory. The current  algebra is:
\begin{align*}
i[\rhd(1), \rhd_m(2)]&=\partial_m\delta(1-2) \\
\EuScript{S}_m&=\rhd\rhd_m\\
\EuScript{U}&=\rhd_m\epsilon^{mn}\partial_n \\
\end{align*}
It fits in the general pattern, by renaming the scalar a dual 2-form $\rhd\to \rhd^{mn}=\epsilon^{mn}\rhd$ and no other changes. 
\item[$\bullet$]D=2
\\Here we just break down the $SL(2)$, toghter with some renaming work:
\begin{align*}
    i[\rhd_{mi}(1),\rhd_{nj}(2)]&=\epsilon_{mnp}C_{ij}\partial^p\delta(1-2) \\
    \to i[\rhd_{m}(1),\rhd^{np}(2)]&=\delta_m^{[n}\partial^{p]}\delta(1-2) \\
    \rhd_{m}&=\rhd_{m1}\\
    \rhd^{mn}&=\epsilon^{mnp}\rhd_{p2}\\
\end{align*}
\item[$\bullet$]D=3
\\
Now we are starting to see some more interesting results. Picking up a direction and decomposing $SL(5)\to GL(4)$, we would break the currents $\rhd$ $10\to 6\oplus 4$, $\EuScript{S}$ constraint (and worldlvolume derivative $\partial$ as its zero-mode) $\bar{5}\to \bar{4}\oplus1$, and $\EuScript{U}$ constraint $5\to 4\oplus 1$:
\begin{align*}
    \rhd_{mn}=P_{mn}+\frac{1}{2}\epsilon_{mnpqr}\partial^pX^qr\to \rhd_{ij}=&P_{ij}+\frac{1}{2}\epsilon_{ijkl}(\partial^kX^l+\partial X^{kl})\\
    \rhd_i=&P_i+\frac{1}{2}\epsilon_{ijkl}\partial^j X^{kl}\\
\end{align*}
\begin{align*}
    \EuScript{S}^m=\epsilon^{mnpqr}\rhd_{np}\rhd_{qr}\to \EuScript{S}^i&=\epsilon^{ijkl}\rhd_{j}\rhd_{kl}\\
    \EuScript{S}&=\epsilon^{ijkl}\rhd_{ij}\rhd_{kl}\\
\end{align*}
\begin{align*}
    \EuScript{U}_m=\rhd_{mn}\partial^n\to \EuScript{U}_i&=\rhd_i \partial +\rhd_{ij}\partial^j\\
    \EuScript{U}&=\rhd_i\partial^i \\
\end{align*}
Now, after applying the selfduality condition on $\rhd_{ij}$, which indicates the relationship 
$P_{ij}=\frac{1}{2}\epsilon_{ijkl}(\partial^kX^l+\partial X^{kl})$, and putting it into $\EuScript{U}_i$, we then consider this component of $\EuScript{U}$ constraint evaluated at the same point:
$$
\EuScript{U}_i=\partial\rhd_i+\epsilon_{ijkl}\partial^j\partial X^{kl}
$$
Thus we can set $\partial=0$ to find a solution, and the algebra, together with the constraints, turns into:
\begin{align*}
    \rhd_{ij}&=P_{ij}+\frac{1}{2}\epsilon_{ijkl}\partial^kX^l \\
    \rhd_i&=P_i+\frac{1}{2}\epsilon_{ijkl}\partial^j X^{kl}\\
    \EuScript{S}^i&=\epsilon^{ijkl}\rhd_{j}\rhd_{kl}\\
    \EuScript{S}&=\epsilon^{ijkl}\rhd_{ij}\rhd_{kl}\\
    \EuScript{U}_i&=\rhd_{ij}\partial^j\\
    \EuScript{U}&=\rhd_i\partial^i \\
\end{align*}
Note that we still keep all the constraints as they could be evaluated at different points.
\item[$\bullet$]D=4
\\From now on, $\EuScript{V}$ constraint would play the key rule. The decomposition of $SO(5,5)$ selfdual current under $GL(6)$ is $16\to 10 \oplus \bar{5} \oplus 1$:
\begin{align*}
    \rhd_\alpha=P_\alpha +\frac{1}{2}\gamma_{\alpha\beta}^i\partial_iX^\beta\to\rhd_{mn}=&P_{mn}+\frac{1}{2}\epsilon_{mnpqr}\partial^r X^{pq}+\frac{1}{2}\partial_{[m}X_{n]}\\
     \rhd^m=&P^m+\frac{1}{2}\partial_nX^{mn}+\frac{1}{2}\partial^mX\\
     \rhd=&P+\frac{1}{2}\partial^mX_m\\
\end{align*}
The $\EuScript{S}$ $10\to 5\oplus \bar{5}$ and $\EuScript{U}$ $\bar{16}\to \bar{10}\oplus5\oplus1$ are correspondingly:
\begin{align*}
    \EuScript{S}_i=\rhd_\alpha\gamma^{\alpha\beta}_i\rhd_\beta \to &\EuScript{S}_m=\rhd^n\rhd_{mn} \\
    &\EuScript{S}^m=\rhd \rhd^m+\epsilon^{mnpqr}\rhd_{np}\rhd_{qr}\\
\end{align*}
\begin{align*}
    \EuScript{U}^\alpha=\gamma^{\alpha\beta}_i\rhd_\beta\partial^i\to \EuScript{U}^{mn}&=\epsilon^{mnpqr}\rhd_{pq}\partial_r+\rhd^{[n}\partial^{m]}\\
    \EuScript{U}_m&=\rhd\partial_m+\rhd_{mn}\partial^n \\
    \EuScript{U}&=\rhd^m\partial_m \\
\end{align*}
At the mean time, $\EuScript{V}$ constraint becomes:
$$
\EuScript{V}=\eta^{ij}\partial_i\partial_{j}=\partial_m\partial^m=0
$$
whose solution, $\partial^m=0$, helps us drop some terms. Now the selfdual current looks like:
\begin{align*}
    \rhd_{mn}&=P_{mn}+\frac{1}{2}\partial_{[m}X_{n]}\\
    \rhd^m&=P^m+\frac{1}{2}\partial_nX^{mn} \\
    \rhd&=P\\
\end{align*}
The scalar current commutes which everything else and thus is decoupled from the algebra and can be dropped by applying the selfduality constraint $\Tilde{\rhd}=P=0$. With this taken into consideration, the non trivial constraints are:
\begin{align*}
    \EuScript{S}_m&=\rhd^n\rhd_{mn} \\
     \EuScript{S}^m&=\epsilon^{mnpqr}\rhd_{np}\rhd_{qr} \\
    \EuScript{U}^{mn}&=\epsilon^{mnpqr}\rhd_{pq}\partial_r \\
    \EuScript{U}&=\rhd^m\partial_m \\
\end{align*}
\item[$\bullet$]D=5
\\First, we take a look at the decomposition $E_6\to SL(6)SL(2)\to GL(6)$, under which the behaviour of current $\rhd$ ($27\to (15,1)\oplus(\bar{6},2)$), $\EuScript{S}$ ($\bar{27}\to (\bar{15},1)\oplus(6,2)$), $\EuScript{U}$ ($78\to (35,1)\oplus (20,2)\oplus(1,3)$), $\EuScript{V}$ ($27\to (15,1)\oplus(\bar{6},2)$) would be respectively:
\begin{align*}
    \rhd_a=P_a+\frac{1}{2}d_{abc}\partial^bX^c\to \rhd_{mn}&=P_{mn}+\frac{1}{2}\epsilon_{mnpqrs}\partial^{pq}X^{rs}+\frac{1}{2}C_{ij}\partial_{[m}^iX_{n]}^j\\
    \rhd^m_i&=P^m_i+\frac{1}{2}C_{ij}(\partial^{mn}X_n^j+\partial_n^jX^{mn})\\
\end{align*}
\begin{align*}
    \EuScript{S}^a=d^{abc}\rhd_b\rhd_c\to \EuScript{S}^{mn}&=\epsilon^{mnpqrs}\rhd_{pq}\rhd_{rs}+C^{ij}\rhd^m_i\rhd^n_j\\
    \EuScript{S}_{mi}&=\rhd^n_i\rhd_{mn}
\end{align*}
\begin{align*}
    \EuScript{U}_A=F_{Ab}{}^c\rhd_c\partial^b\to \EuScript{U}_M&=f_{Mmn}{}^{pq}\rhd_{pq}\partial^{mn}+f_{Mm}{}^n\rhd^n_i\partial_m^i \\
    \EuScript{U}_{mnp,i}&=\rhd_{[mn}\partial_{p]i}+\epsilon_{mnpqrs}\rhd^q_i\partial^{rs} \\
    \EuScript{U}_{ij}&=\rhd^m_{(i}\partial_{|m|j)} \\
\end{align*}
\begin{align*}
    \EuScript{V}_a=d_{abc}\partial^b\partial^c\to \EuScript{V}_{mn}&=\epsilon_{mnpqrs}\partial^{pq}\partial^{rs}+C_{ij}\partial_{m}^i\partial_n^j \\
    \EuScript{V}^m_i&=\partial^{mn}\partial_{ni}\\
\end{align*}
Solving $\EuScript{V}$ constraint first would further break the $SL(2)$ and we get $\partial^{mn}=\partial_{m2}=0$. Renaming $\partial_{m1}=\partial_m^2=\partial_m$ and then putting it back to the currents, we would have:
\begin{align*}
    \rhd_{mn}&=P_{mn}+\frac{1}{2}\partial_{[m}X^1_{n]} \\
    \rhd^m_1&=P^m_1+\frac{1}{2}\partial_nX^{mn} \\
    \rhd_2^m&=P_2^m \\
\end{align*}
Again, $\rhd^m_2$ is decoupled from the rest of the algebra and can be dropped out by applying the selfduality condition $\Tilde{\rhd}^m_2=P^m_2=0$. Thus the $\EuScript{S}$ and $\EuScript{U}$ constraints for M theory in this case would be:
\begin{align*}
    \EuScript{S}^{mn}&=\epsilon^{mnpqrs}\rhd_{pq}\rhd_{rs} \\
    \EuScript{S}_m&=\rhd^n\rhd_{mn}\\
    \EuScript{U}_{mnp}&=\rhd_{[mn}\partial_{p]} \\
    \EuScript{U}&=\rhd^m\partial_m \\
\end{align*}
\item[$\bullet$]D=6
\\Although the technical detail of $E_7\to GL(7)$ branching is messy, the idea in general is the same. We have:
\begin{align*}
    \rhd \& X(56)\to& 7_{-6}\oplus 21_{2}\oplus \bar{21}_{-2}\oplus\bar{7}_6\\
    \partial\&\EuScript{S}\&\EuScript{V}(133)\to& 48_0\oplus 1_0\oplus 7_8\oplus\bar{7}_{-8}\oplus 35_{-4}\oplus \bar{35}_4\\
    \EuScript{U}(912)\to& 28_2\oplus7_{-6}\oplus1_{-14}\oplus140_{-6}\oplus224_2\oplus21_2\oplus35_{10}\\
    \oplus&\bar{28}_{-2}\oplus\bar{7}_6\oplus\bar{1}_{14}\oplus\bar{140}_6\oplus\bar{224}_{-2}\oplus\bar{21}_{-2}\oplus\bar{35}_{-10}\\
\end{align*}
Solving the $\EuScript{V}$ constraint $133\otimes 133\to 133$ would help us select the proper worldvolume directions we want to keep, which could then tell us how to simplify the rest stuffs:
\begin{align*}
    \EuScript{V}_{48\oplus1}&=\partial_{7}\otimes\partial_{\bar{7}}+\partial_{35}\otimes\partial_{\bar{35}} \\
    \EuScript{V}_7&=\partial_{48\oplus1}\otimes\partial_7\\
    \EuScript{V}_{\bar{7}}&=\partial_{48\oplus1}\otimes\partial_{\bar{7}}\\
    \EuScript{V}_{35}&=\partial_{48\oplus 1}\otimes\partial_{35}+\partial_{\bar{7}}\otimes\partial_{\bar{35}} \\
    \EuScript{V}_{\bar{35}}&=\partial_{48\oplus 1}\otimes\partial_{\bar{35}}+\partial_{7}\otimes\partial_{35} \\
\end{align*}
Choosing $\partial_7=\partial_m$ \& $\partial_{\bar{35}}=\partial^{[mnp]}$ (or $\partial_{\bar 7}=\partial^m$ \& $\partial_{35}=\partial_{[mnp]}$) being non-zero may satisfy all the requirements.
Therefore, the self-dual currents are now:
\begin{align*}
    \rhd_7=&\rhd_m=P_m \\
    \rhd_{21}=&\rhd_{mn}=P_{mn}+\frac{1}{2}\partial_{[m}X_{n]}+\frac{1}{2}\epsilon_{mnpqrst}\partial^{pqr}X^{st}\\
    \rhd_{\bar{21}}=&\rhd^{mn}=P^{mn} \\
    \rhd_{\bar7}=&\rhd^m=P^m+\frac{1}{2}\partial_nX^{mn}+\frac{1}{2}\partial^{mnp}X_{np} \\
\end{align*}
Dropping $\rhd_m$, which commutes with everything else, by applying selfduality condition first, then looking at the $224\oplus21$ component of $\EuScript{U}$ constraint, which contains only $\rhd^{mn}$ and $\partial^{mnp}$ now, we know that setting $\partial_{\bar{35}}=0$ would be a solution. Thus we also have $\rhd^{mn}$ being decoupled, as we can remove $\partial^{mnp}X_{np}$ from $\rhd^m$. We drop it out as well. Finally, we summarize all the remaining pieces:
\begin{align*}
    \rhd^m&=P^m+\frac{1}{2}\partial_nX^{mn} \\
    \rhd_{mn}&=P_{mn}+\frac{1}{2}\partial_{[m}X_{n]} \\
    \EuScript{S}_m&=\rhd_{mn}\rhd^n \\
    \EuScript{S}^{mnp}&=\epsilon^{mnpqrst}\rhd_{qr}\rhd_{st} \\
    \EuScript{U}&=\rhd^m\partial_m \\
    \EuScript{U}_{mnp}&=\rhd_{[mn}\partial_{p]} \\
\end{align*}
\end{enumerate}
\section{Conclusion}
Starting with a action of a fundamental brane living in $GL(D+2)/SO(D,2)$, one can reveal the M theory algebra which manifests the $GL(D+1)$ group. The contents of background fields, together with their gauge transformation rules, being read out following the standard procedures, are consistent with the result from the usual supergravity field theoretic approach.  This approach of M theory could also be treated as a subset of the F theory, which is covariant under the whole $E_{D+1}$ U group, by solving $\EuScript{V}$ sectioning explicitly.
\\

The M theory formulated in this way could also be treated as a toy model or prototype of first quantized F theory, as it has simpler group structure yet still keeps all the key features: 
\begin{enumerate}
    \item[$\bullet$] Gauge structure of the coordinate mapping from worldvolume to spacetime
    \item[$\bullet$] Correlation between worldvolume and spacetime vielbein
    \item[$\bullet$] Enlarged symmetry in Lagrangian description
\end{enumerate}

By taking a more carefully look into it we may be illuminated and come up with a better understanding on some open problems we are facing in F theory construction. Thus there are interesting topics worth discussion in the future:
\begin{enumerate}
    \item[$\bullet$] M theory Lagrangian with curved background
    \item[$\bullet$] Supersymmetric M theory action
    \item[$\bullet$] Detailed behaviour of the spectrum under some particular background
\end{enumerate}
\section*{Acknowledgements}
This work is supported by NSF grant PHY-1915093. We thank Machiko Hatsuda for helpful discussions.  
\appendix
\section{Notations \& Indices}
Here we list the meaning of corresponding notations and indices in the main context:
\begin{center}
\begin{tabular}{cc}
X & spacetime coordinate \\
P & spacetime momentum \\
$\sigma$ & worldvolume coordinate \\
$\partial$ & worldvolume derivative \\
$\EuScript{S}$ & Virasoro constraint \\
$\EuScript{U}$ & Gaussian constraint \\
$\EuScript{V}$ & worldvolume sectioning constraint \\
$M$,$N$... & abstract spacetime index \\
$\EuScript{I}$,$\EuScript{J}$... & abstract worldvolume index \\
$m$,$n$... or $i$,$j$... & SL(n) index\\
$\lambda$ & abstract gauge index\\
$\alpha$,$\beta$... & bosonic spinor index\\
\end{tabular}
\end{center}
\bibliography{Warren,others}
\bibliographystyle{hephys}
\end{document}